\documentclass[superscriptaddress,groupedaddress,nofootnoteinbib,11pt]{article}  
\pdfoutput=1

\usepackage{graphicx}  
\usepackage{dcolumn}   
\usepackage{bm}        
\usepackage{slashed}        
\usepackage{float}
\usepackage{jcappub}
\usepackage{color}

\usepackage{graphicx}
\usepackage{dcolumn}
\usepackage{bm}
\usepackage{amssymb}
\usepackage{amsmath}
\usepackage{sectsty}
\usepackage{colortbl}
\usepackage{latexsym}
\usepackage{float}
\usepackage{ifthen}
\usepackage{enumerate}
\usepackage{url}
\usepackage{jcappub}
    \usepackage{picinpar}
    \usepackage{colortbl}
\usepackage{multirow}
	\usepackage{float}
	      \usepackage{setspace}
\usepackage{array}
\usepackage{bm}
\usepackage{amsopn}

\usepackage{caption}
\usepackage{subcaption}

\usepackage{booktabs}
\usepackage[table]{xcolor}

\usepackage{slashed}

\begin{document}

\title{Large Slow Roll Parameters in Single Field Inflation}

\author[a]{Jessica L. Cook,}
\author[a,b]{Lawrence M. Krauss}
\affiliation[a]{Department of Physics and School of Earth and Space Exploration \\ Arizona State University, Tempe, AZ 85827-1404}
\affiliation[b]{Research School of Astronomy and Astrophysics, Mt. Stromlo Observatory, \\ Australian National University, Canberra, Australia 2611}

\emailAdd{jlcook14@asu.edu}
\emailAdd{krauss@asu.edu}
\date{\today}
\abstract{

We initially consider two simple situations where inflationary slow roll parameters are large and modes no longer freeze out shortly after exiting the horizon, treating both cases analytically. By modes, we refer to the comoving curvature perturbation $R$. We then consider applications to transient phases where the slow roll parameters can become large, especially in the context of the common `fast-roll' inflation frequently used as a mechanism to explain the anomalously low scalar power at low $l$ in the CMB. These transient cases we treat numerically. We find when $\epsilon$, the first slow roll parameter, and only $\epsilon$ is large, modes decay outside the horizon, and when $\delta$, the second slow roll parameter, is large, modes grow outside the horizon.  When multiple slow roll parameters are large the behavior in general is more complicated, but we nevertheless show in the 'fast-roll' inflation case, modes grow outside the horizon.

 }
\maketitle

\section{Introduction}

The simplest inflaton models assume that slow roll is valid during the whole period from when current CMB scales first exited the horizon up to the time just before inflation ends. There are reasons one might suppose this isn't the case. Obviously the slow roll parameters must become large as inflation is ending, but since the modes we do observe were far outside the horizon when inflation ended, we can presumably ignore this complication. We know the slow roll parameters couldn't have been large during the whole of inflation because either:  (a) inflation wouldn't have lasted long enough, or (b) the observed $n_s$ would look substantially different from what has been observed ($n_s \ll 1$ if $\epsilon$ is close to 1 during much of inflation, or $n_s \approx 1$ in the ultra flat $\delta \approx 3$ case. More on these below.)   It would be phenomenologically interesting, however, if the slow roll parameters became large for transient periods during which scales that are currently observable today first left the horizon.  For example, something along the lines of \cite{Takahashi:2013tj} where the potential is comprised of a small amplitude, high frequency trig function superimposed on a larger amplitude smaller frequency function would naturally lead to oscillations in the slow roll parameters. One can get oscillations with large amplitudes in the higher order slow roll parameters like $\delta$ and $\xi^2$ while $\epsilon$ remains small and inflation last long enough. In \cite{Takahashi:2013tj} this was used as a possible explanation for signs of negative running in SPT data \cite{Hou:2012xq}, signs that haven't been replicated in Planck data \cite{Ade:2015lrj}. 

A second possibility is that the slow roll parameters started out large at the time the largest CMB scales were exiting the horizon, which might be possible if these scales corresponded to the beginning of inflation. There might even be a hint for this found in the CMB in the anomalously low scalar power at low $l$. There are various reasons one might suppose inflation didn't last much longer than necessary to solve the horizon, flatness problem, etc. For example, one might suppose it's unnatural for a field to end up too far offset from its minimum. If inflation were preceded by a symmetry breaking event, one wouldn't assume the separation between the old and new minima to be arbitrarily large, especially if one assumes the breaking happened below the Planck scale. In another common picture where inflation started from a random quantum fluctuation, small quantum fluctuations should be exponentially more common than large ones, and many small fluctuations stacking up in the same direction are also disfavored, leading to the inflaton likely not ending up much further offset from its minimum than necessary.

The significance of the low $l$ power anomaly is around $2.5 - 3 \sigma$ \cite{Ade:2013kta}. The anomaly was first discovered in the COBE data \cite{Bennett:1996ce}, and has since persisted in the WMAP and Planck data \cite{Bennett:2003bz, Spergel:2003cb,Ade:2013kta, Ade:2015lrj}. It might just be an effect of cosmic variance, {for example see \cite{White:1993jr}}.  Moreover, since the CMB comprises so many independent measurements, it would be surprising if none of those measurements ended up deviating in a significant way from the theoretical expectation \cite{Contaldi:2003zv}. Alternatively, it could be a hint of non-slow roll behavior at those scales, possibly providing insight into the beginning of inflation. If large tensors are observed in the future, this will increase the significance of the anomaly. The reason is that tensors as well as scalars contribute to $\langle TT \rangle$ at low $l$ before around $l \approx 100$ after which tensors drop off substantially. Therefore, the fact that $\langle TT \rangle$ at low $l$ is already smaller than expected based on the rest of the power spectrum, observing large tensors and inferring their contributions to $\langle TT \rangle$ would mean the scalar contribution would have to be even more suppressed at those $l$ than previously thought. 

For these reasons, it's interesting to see how large slow roll parameters affect the power spectrum. In particular, when slow roll no longer holds, modes no longer necessarily freeze out shortly after exiting the horizon \cite{Leach:2000yw,Leach:2001zf,Jain:2007au,Jain:2008dw,Jain:2009pm}. By modes, we refer to the spatial curvature pertrubation $R$ evaluated in comoving gauge, though one could just as well consider $\zeta$, the gravitational potential evaluated in the gauge without density perturbations. This has a direct impact on the power spectrum which can be defined by: $P = \frac{k^3}{2 \pi^2} \lim_{\frac{k}{aH} \rightarrow 0} |R_k|^2$. $R$ (or $\zeta$) typically asymptote to a constant shortly after a $k$ mode leaves the horizon. This is the case when the perturbations are largely adiabatic, with an irrelevant entropy perturbation. However, there are situations were modes don't immediately freeze out, even for single field inflation. This can be understood as due to an entropy perturbation staying relevant compared to the adiabatic perturbation. We find that the gauge invariant metric perturbations can actually grow (or shrink) exponentially outside the horizon if they exit the horizon when the slow roll parameters are large. Therefore, for these modes, it is important to evaluate them after slow roll is reached and not at horizon crossing, or one will underestimate (or overestimate) the amplitude of the predicted power spectrum for these modes. 


The papers \cite{Leach:2000yw,Leach:2001zf} gave a particularly useful way of understanding why modes grow in the $\delta \approx 3$ case. Note the equation of motion for the comoving curvature perturbation $R$:

 \begin{align}
R_{\tau \tau} + 2 \frac{z_{\tau}}{z} R_{\tau} + k^2 R =0
\end{align}

\noindent where $\tau$ is conformal time and $z = \frac{a \phi_{\tau}}{\mathcal{H}}$. The $ 2 \frac{z_{\tau}}{z} R_{\tau} $ acts as a friction term, and when it becomes negative, $R$ grows on super horizon scales. $\frac{z_{\tau}}{z}$ can be written as $ 1 - \delta + \epsilon$, and note $\epsilon$ is always less than 1 if the universe is inflating, so if during inflation, $\delta$ is positive and larger than $1 + \epsilon$, then the friction term is negative.

\cite{Jain:2007au,Jain:2008dw,Jain:2009pm} consider connections of superhorizon evolution to the low power at low $l$ anomaly. \cite{Jain:2007au} considers superhorizon evolution for DBI inflation, and \cite{Jain:2008dw,Jain:2009pm} considers potentials which lead to temporary breaks in inflation leading to dips followed by enhancement of the power spectrum. They show that if such a dip in the power spectrum aligned with the lowest $l$ scales, this helps alleviate the anomaly. 



We will need the slow roll parameters written as derivatives of the Hubble parameter since the common definition in terms of derivatives of only the potential assumes slow roll in the derivation and also doesn't account for different choices of initial conditions. We use the standard definition for $\epsilon$ 

 \begin{align}
\epsilon = 2 M_P^2 \frac{H_{\phi}^2}{H^2}
\end{align}

\noindent (where subscript $\phi = \frac{d}{d \phi}$, $H$ is the Hubble parameter, and $M_P$, the reduced Planck mass) and then follow the prescription in \cite{Liddle:1994dx} for the higher slow roll parameters. These are defined such that the order $n$ parameter is given by  

 \begin{align}
^n \beta = 2 M_P^2 \left(\frac{H_{\phi}^{n-1} \, (\frac{d}{ d \phi })^{n+1} H}{H^n} \right)^{\frac{1}{n}} 
\end{align}

\noindent where $n=1$ gives $\delta$ and $n = 2$ gives $\xi$ and so on. While in principle there is an infinite number of slow roll parameters incorporating higher order derivatives of the Hubble parameter, only the first three, $\epsilon$, $\delta$ and $\xi^2$, appear in the general Mukhanov-Sasaki (MS) equation, and are the only ones we will use here. The above prescription gives for $\delta$ and $\xi^2$:

 \begin{align}
\delta = 2 M_P^2 \frac{H_{\phi \phi}}{H}
\end{align}

\noindent and

 \begin{align}
\xi^2 = 4 M_P^4 \frac{H_{\phi} H_{\phi \phi \phi}}{H^2} \, .
\end{align}

It will be convenient to work with these equations with time derivatives instead, where the slow roll parameters can be rewritten:

 \begin{align}\label{eq5}
\epsilon =  \frac{\phi_N^2}{2 M_P^2}  \hspace{6mm}  \delta = \frac{\phi_{NN}}{\phi_N} + \frac{\phi_N^2}{2 M_P^2}  \hspace{6mm} \xi^2 = \frac{3}{2 M_P^2} \phi_{NN} \phi_N + \frac{1}{4 M_P^4} \phi_N^4 + \frac{\phi_{NNN}}{\phi_N} - \frac{\phi_{NN}^2}{\phi_N^2}
\end{align}

\noindent in efolding time ($N$), or in conformal time ($\tau$):

 \begin{align}
\epsilon = \frac{1}{2 M_P^2} \frac{\phi_{\tau}^2}{\mathcal{H}^2} \hspace{6mm} \delta = 1 - \frac{\phi_{\tau \tau}}{\mathcal{H} \phi_{\tau}} \hspace{6mm}  \xi^2 = - \frac{\phi_{\tau \tau}}{\mathcal{H} \phi_{\tau}} + \frac{1}{\mathcal{H}^2} \frac{\phi_{\tau \tau \tau}}{\phi_{\tau}} - \frac{1}{\mathcal{H}^2} \frac{\phi_{\tau \tau}^2}{\phi_{\tau}^2} + \frac{\phi_{\tau}^2}{2 M_P^2 \mathcal{H}^2} \, ,
\end{align} 



\noindent where $\mathcal{H} = \frac{a_{\tau}}{a}$. These can then be applied to the Mukhanov-Sasaki equation. The familiar form of the equation: $u_{\tau \tau} + u(k^2 - \frac{z_{\tau \tau}}{z})=0$ with $z = \frac{a \phi_{\tau}}{\mathcal{H}}$ is general and doesn't assume slow roll. It is at the stage of expanding out $z_{\tau}$ and $z_{\tau \tau}$ that one generally drops higher order slow roll terms, but one can easily leave them in to obtain:

 \begin{align}
u_{\tau \tau} + u \left(k^2 - \frac{\phi_{\tau \tau \tau}}{\phi_{\tau}} - 2 \frac{\phi_{\tau \tau} \phi_{\tau}}{M_P^2 \mathcal{H}} + \frac{\phi_{\tau}^2}{2 M_P^2} - \frac{\phi_{\tau}^4}{2 M_P^4 \mathcal{H}^2} \right) = 0  \, .
\end{align}

\noindent The derivative of $\phi$ terms can then be traded for slow roll parameters:

\begin{align}\label{eq1}
u_{\tau \tau} + u(k^2 - \mathcal{H}^2 (2 - 3 \delta + 2 \epsilon + \xi^2 + \delta^2 - 4 \epsilon \delta + 2 \epsilon^2)) = 0  \, ,
\end{align}

\noindent or equivalently in efolding time:

\begin{align}
u_{NN} + u_N (\epsilon -1) + u \left(\frac{k^2}{H^2} e^{2N} - 2 + 3 \delta - 2 \epsilon - \xi^2 - \delta^2 + 4 \epsilon \delta - 2 \epsilon^2 \right) = 0 \, .
\end{align}

In section 2 we consider a simple case where $\epsilon \approx 1$ and stays $\approx 1$ for the duration of inflation, and show there is a delayed freeze out effect. In section 3 we explore a simple case of $\delta \approx 3$ for the duration of inflation, and determine how modes actually grow outside the horizon. This case has been considered before, \cite{Tsamis:2003px, Kinney:2005vj, Tzirakis:2007bf, Namjoo:2012aa, Martin:2012pe, Motohashi:2014ppa, Mooij:2015yka} but is included here for completeness, and for relevance to transient situations where the slow roll parameters become large. In section 4 we consider numerical studies of modified freeze out with transient periods of large fast roll parameters. In section 5 we consider a simple analytic approximation of one of the transient phases that well approximates the final power spectrum. In section 6 we present our conclusions. 

\section{$\epsilon$ Large}





 Suppose $\epsilon$ is close to but slightly less than one such that the universe still inflates.  Note that $\epsilon = \frac{\phi_N^2}{2 M_P^2}$, using efolding time, $N$.  Thus for $\epsilon \approx 1$,  $\phi_N \approx \sqrt{2} M_P$. One can always start with initial conditions with $\epsilon$ this large, but generally this will lead to $\phi_N$ decreasing exponentially until a slow roll solution is reached. The quintessential example for $\epsilon$ large is to inflate on a potential such that the asymptotic solution has $\phi_N \approx \sqrt{2} M_P$. 
 
The equation of motion of $\phi$ using efolding time is given by:
 
  \begin{align}
\phi_{NN} + (-3 + \epsilon) \phi_N + \frac{1}{H^2} \partial_{\phi} V = 0 \, .
 \end{align}
 
\noindent For an example where $\epsilon \approx 1$ is maintained for a long period, we need to maintain $\phi_N \approx \sqrt{2} M_P$, and this requires a solution with $\phi_{NN} \approx 0$. Since $\epsilon \approx 1$, the equation of motion becomes:
 
   \begin{align}
 2 H^2 \phi_N =  \partial_{\phi} V 
 \end{align}
 
\noindent Next we use the Friedmann equation in efolding time: $H^2 = \frac{V}{3 M_P^2 - \frac{\phi_N^2}{2}}$ and plug this into the equation of motion to give: 
 
      \begin{align}
V =  \frac{M_P}{\sqrt{2}} \partial_{\phi} V  \, .
 \end{align}
 
\noindent So a potential that maintains $\epsilon \approx 1$ requires  $V \propto \partial_{\phi} V $.  This is obviously satisfied by exponentials,  so suppose 
 
       \begin{align}
V = \Lambda^4 e^{\frac{\phi}{f}} \, .
 \end{align}
 
\noindent  Plugging this into $V =  \frac{M_P}{\sqrt{2}} \partial_{\phi} V $, we find

  \begin{align}
f = \frac{M_P}{\sqrt{2}} \, .
 \end{align}

Thus if $V = \Lambda^4 e^{\frac{\phi}{f}}$ with $f = \frac{M_P}{\sqrt{2}}$, this allows for inflation with $\epsilon \approx 1$. In order that $\epsilon$ is actually a little less than $1$, $f$ should be slightly larger that $\frac{M_P}{\sqrt{2}}$. 

We consider the other slow roll parameters for this potential (equation \ref{eq5}). We find  $\frac{\phi_{NN}}{\phi_N}  \approx 0$, but $\delta \approx 1$ since $\delta = \frac{\phi_{NN}}{\phi_N} + \epsilon$. Similarly we find $\frac{\phi_{NNN}}{\phi_N} \approx 0$, but $\xi^2 \approx 1$ since $\xi^2 = \frac{\phi_{NNN}}{\phi_N} + 5 \epsilon \delta - 3 \epsilon^2 - \delta^2$. Note that it is the large $\epsilon$ value that makes $\delta$ and $\xi^2 \approx 1$. We will use these to solve the full Mukhanov-Sasaki equation without slow roll approximations (equation \ref{eq1}).  Let $\epsilon = 1 - \alpha$, so $\alpha$ is positive and $\alpha \ll 1$.  The MS equation becomes:

 \begin{align}
u_{\tau \tau} + u \left( k^2 - \mathcal{H}^2 (1+ \alpha ) \right) = 0
 \end{align}
 
 Since $\phi_{NN} \approx 0$, $\epsilon \approx$ constant, then $\alpha \approx$ constant. We plug in for $\mathcal{H}$ using the definition for $\epsilon$ in conformal time: $\frac{\mathcal{H}_{\tau}}{\mathcal{H}^2}= 1 - \epsilon$:
 
  \begin{align}
\int \frac{1}{\mathcal{H}^2} d \mathcal{H} = \int \alpha d \tau \, .
 \end{align}
 
 \noindent After choosing to define $\tau$ such that the integration constant is 0, this integrates to: 
 
 \begin{align}
\mathcal{H} = \frac{1}{- \alpha \tau } \, .
 \end{align}
 
 \noindent Then the MS equation becomes:

 \begin{align}
u_{\tau \tau} + u \left( k^2 - \frac{1}{\alpha^2 \tau^2} (1+ \alpha ) \right) = 0 \, ,
 \end{align}

 \noindent which can be solved:

 \begin{align}
u = c_1 \sqrt{- \tau} H^{(1)}_{\frac{1}{\alpha} + \frac{1}{2}} ( - k \tau) + c_2 \sqrt{- \tau} H^{(2)}_{\frac{1}{\alpha} + \frac{1}{2}} ( - k \tau) \, .
 \end{align}
 
 \noindent We recognize this is the standard slow roll result except with $\nu \rightarrow \frac{1}{\alpha} + \frac{1}{2}$, and so matching onto Bunch-Davies (BD) initial conditions gives:
 
  \begin{align}
u =  \frac{\sqrt{\pi}}{2} \sqrt{- \tau} H^{(1)}_{\frac{1}{\alpha} + \frac{1}{2}} ( - k \tau)  \, .
 \end{align}
 
 \noindent  This can be rewritten using $\tau = \frac{-1}{\alpha a H}$:
 
  \begin{align}
u =  \frac{\sqrt{\pi}}{2} \sqrt{\frac{1}{\alpha a H}} H^{(1)}_{\frac{1}{\alpha} + \frac{1}{2}} \left( \frac{k}{\alpha a H} \right) \, .
 \end{align} 
 
Normally during slow roll $\nu \approx \frac{3}{2}$, and there are only small deviations from $\frac{3}{2}$ which produce a tilt in the power spectrum. Here we have $\nu \approx \frac{1}{\alpha}$ where $\alpha \ll 1$, which will create an extremely red power spectrum. The closer $\epsilon$ gets to one, the more tilted the power spectrum becomes.

Also note that, as always, modes exit the horizon when $k = aH$ but now the long wavelength limit of the mode function only becomes valid when $k \ll \alpha a H$, and since $\alpha \ll 1$, this occurs well after a mode has exited the horizon.  This means the amplitude for modes will continue to decay for a long period after they have exited the horizon, but freeze out will eventually occur when $k \ll \alpha a H$. The closer $\epsilon$ is to one, the longer it takes for freeze-out to eventually happen. 
 
We next calculate the power spectrum approximated for modes with $k \ll \alpha a H$.  Using the gauge invariant perturbation $|R|^2 = \frac{|u|^2}{a^2 \phi_N^2}$ and $P_{\zeta} = \frac{k^3}{2 \pi^2} \lim_{\frac{k}{\alpha a H} \rightarrow 0} |R|^2 $:
 
  \begin{align}
P_{\zeta} = \frac{ 2^{2 \nu} \Gamma^2(\nu)}{8 \pi^3  \phi_N^2}(\alpha  H)^2  \left( \frac{\alpha a H}{k } \right)^{\frac{2}{\alpha} -2} \,.
 \end{align} 
 
 \noindent We find $P_{\zeta} \propto (\frac{1}{k})^{\mbox{large\, number}}$ giving an extremely red power spectrum. 
 
To show that the power spectrum does eventually freeze out, we plug $H$ and $\phi_N$ into $P_{\zeta}$ to better see the $N$ dependence. Using the equation of motion above:
 
   \begin{align}
\phi_N = \frac{M_P^2}{f} \, .
 \end{align}

\noindent  Let $N_0$ define the initial value for $N$ with $N$ counting down towards 0 at the end of inflation. Then integrating the above gives:

\begin{align}
\phi =  \frac{M_P^2}{f} (N - N_0) + \phi_0 \, .
 \end{align}
 
 \noindent We use the Friedmann equation to obtain: 
 
   \begin{align}
H = \frac{\Lambda^2}{\sqrt{3 M_P^2 - \frac{1}{2} \frac{M_P^4}{f^2}}} e^{\frac{1}{2f}(\phi_0 + \frac{M_P^2}{f}(N-N_0))} \, .
 \end{align}

 \noindent We will focus only on the time dependent part:
 
   \begin{align}
P_{\zeta} \propto e^{\frac{M_P^2}{f^2} N} \left( e^{ N  \left(\frac{M_P^2}{2 f^2} -1\right) } \right)^{\frac{2}{\alpha} -2} \, .
 \end{align}
 
Next, we determine $f$ as a function of $\alpha$. We use $\epsilon = \frac{\phi_N^2}{2 M_P^2}$ and $\phi_N = \frac{M_P^2}{f}$ which gives

   \begin{align}
f^2 = \frac{M_P^2}{2 (1 - \alpha)} \, .
 \end{align}
 
 \noindent This yields the power spectrum:
 
  \begin{align}
P_{\zeta} \propto e^{2 N (1 - \alpha)} \left(e^{N ((1 - \alpha) - 1)} \right)^{\frac{2}{\alpha} -2} \, .
 \end{align} 
 
 \noindent We  find the $N$ dependence cancels out. So, as expected, the modes do freeze out, but not until $k \ll \alpha a H$.

One can then determine how much longer it takes for modes to freeze out after they exit the horizon, or more specifically how many efolds pass between the period when $k = a H$ and when $k = \alpha a H$. 

 \begin{align}
e^{-N_1} \frac{\Lambda^2}{\sqrt{3 M_P^2 - \frac{1}{2} \frac{M_P^4}{f^2}}} e^{\frac{1}{2f} \left( \phi_0 + \frac{M_P^2}{f}(N_1 - N_0) \right)} = \alpha e^{-N_2} \frac{\Lambda^2}{\sqrt{3 M_P^2 - \frac{1}{2} \frac{M_P^4}{f^2}}} e^{\frac{1}{2f} \left( \phi_0 + \frac{M_P^2}{f}(N_2 - N_0) \right)}
 \end{align}
 
 \noindent Using $f^2 = \frac{M_P^2}{2(1 - \alpha)}$ we find:
 
  \begin{align}
\Delta N = - \frac{\ln \alpha}{\alpha}  \, .
 \end{align}
  
  So for example, for $\epsilon = 0.95$, this corresponds to $\Delta N = 60$. The closer $\epsilon$ is to one, the longer the freeze out time. This means when inflation ends, there will be modes which haven't reached the long wavelength limit, and these modes will have more power than they would otherwise have. 

 

Of course this model produces an entirely wrong $n_s$ and can't describe our actual power spectrum. As we shall explore, more realistic models could contain transient periods where $\epsilon$ becomes large, which is almost certainly true at least at the end of inflation but possibly elsewhere too, and this model can give useful insights for those cases.  

We find this behavior numerically in the arctan potential example below which passes through a period with $\epsilon \approx 1$. There, as we shall see, the behavior is more complicated because as $\epsilon$ returns to the slow roll value, some modes actually grow outside the horizon rather than decay, leading to oscillations in the power spectrum. Solving analytically in these scenarios is very difficult because not only are the slow roll parameters large, their derivatives are large, and one can no longer approximate them as constant in the MS equation. 



\section{$\epsilon$ Small but $\eta$ Large }

When the $V_{\phi}$ term in the equation of motion during inflation is subdominant, one enters a regime where $\delta \approx 3$ until enough kinetic energy is redshifted that $V_{\phi}$ starts to balance the other terms. Then $\delta$ will transition to $\delta \ll 1$ and slow roll will be reached. If one also starts with $\epsilon \approx 1$, and the potential isn't as steep as in the last example, the kinetic energy will drop off, and there will frequently be an in-between regime, where $\epsilon$ starts to drop off, but $\delta \approx 3$ until enough kinetic energy is lost that $\delta$ too becomes small and slow roll is reached. This case will be seen in the $\phi^2$ and $R^2$ examples in section 4. The flatter the potential, the longer the $\delta \approx 3$ regime. We start from  the equation of motion in efolding time:

  \begin{align}
\phi_{NN} + (-3 + \epsilon) \phi_N + \frac{1}{H^2} \partial_{\phi} V = 0 \, .
 \end{align}
 
 If $\epsilon$ has already become small, or was never large to begin with, but the field has enough kinetic energy such that $\partial_{\phi} V$ is much smaller than the other terms, we find:

  \begin{align}\label{eq2}
\phi_{NN} = 3  \phi_N  \, .
 \end{align}
 
Note from equation \ref{eq5} $\delta = \frac{\phi_{NN}}{\phi_N} +\epsilon$. After $\epsilon$ has become small, then $\delta = \frac{\phi_{NN}}{\phi_N}$ so that $\delta = 3$.

The flatter the potential, the longer the $V_{\phi}$ term will stay irrelevant and $\delta =3$. So the quintessential example of trying to keep $\delta$ large would be a perfectly flat potential, $V_{\phi}=0$.

This has been worked out in the past \cite{Tsamis:2003px, Kinney:2005vj, Tzirakis:2007bf, Namjoo:2012aa, Martin:2012pe, Motohashi:2014ppa, Mooij:2015yka}. It has been found that freeze out never occurs for the duration of inflation and modes actually grow exponentially outside the horizon. In spite of this, the final power spectrum is perfectly flat, without features. 

One can derive this by solving the background/ classical equations, including first the Friedmann equation:

 \begin{align}
H = \frac{\sqrt{V} }{\sqrt{3 M_P^2 - \frac{\phi_N^2}{2} }  } \, .
\end{align}

In the regime where $\epsilon$ is small, $\frac{\phi_N^2}{2} \ll 3 M_P^2$, and with a potential where $V$ is constant: 

 \begin{align}
H \approx \sqrt{ \frac{V_0 }{3 M_P^2 } } \, ,
\end{align}

\noindent the Hubble parameter is also essentially constant.  Then we solve the the equation of motion for $\phi$, equation \ref{eq2}, which gives:

 \begin{align}
\phi = \phi_0 +  \frac{\phi_{N\, 0}}{3} \left(e^{3(N - N_0)} -1\right) \, .
\end{align}

Next we solve the full Mukhanov-Sasaki equation, equation \ref{eq1}, where we have both $\epsilon$, $\xi^2 \approx 0$, and $\delta=3$ is constant:

\begin{align}
u_{\tau \tau} + u (k^2 - 2 \mathcal{H}^2  ) = 0 \, .
\end{align}

\noindent We choose initial conditions for $\tau$ such that $\mathcal{H}  = - \frac{1}{\tau}$:

 \begin{align}
u_{\tau \tau} + u (k^2 - \frac{2}{\tau^2} ) = 0 \, .
\end{align}

\noindent This gives the normal slow roll mode function equation which has solution:

 \begin{align}
u = \frac{\sqrt{\pi}}{2} \sqrt{- \tau} H^{(1)}_{\frac{3}{2}}(- k \tau) \, .
\end{align}

So a potential with $\delta = 3$ actually generates the same mode functions as a power spectrum with $\delta =0$, assuming $\epsilon$ and $\xi^2$  $\approx 0$. However, the evolution of the metric perturbations outside the horizon is very different. We use the gauge invariant metric perturbation $|R|^2 = \frac{|u|^2}{a^2 \phi_N^2}$ and $P_{\zeta} = \frac{k^3}{2 \pi^2} \lim_{\frac{k}{ a H} \rightarrow 0} |R|^2 $. The difference comes from the behavior of the $\phi_N$ which is approximately constant during slow roll, but is decaying exponentially when $\delta =3$. This means the behavior of $|R|^2$ for a particular $k$ mode grows exponentially outside the horizon since $|R|^2 \propto \frac{1}{\phi_N^2}$. The power spectrum is:

 \begin{align}
P_{\zeta} = \frac{2^{2 \nu} \Gamma^2(\nu)}{8 \pi^3} \frac{H^2}{\phi_N^2} \, ,
\end{align}

\noindent which is the form of the slow roll power spectrum except all the $k$ dependence of the power spectrum drops out. 

To explore the time dependence, we insert values of $H$ and $\phi_N$. $H$ is approximately constant, but $\phi_N =\phi_{N\, 0} e^{3(N - N_0)}$. 

 \begin{align}
P_{\zeta} = \frac{2^{2 \nu} \Gamma^2(\nu)}{8 \pi^3} \frac{H_0^2}{\phi_{N\, 0}^2 e^{6(N- N_0)}}
\end{align}

Thus $P_{\zeta} \propto e^{- 6 N}$.   Since we are using conventions where $N$ counts down towards the end of inflation, this means the power spectrum is growing exponentially in time. Note that typically inflation doesn't end in this scenario so that one imagines some coupling with another field forces inflation to end at some particular point. Then the power spectrum should freeze out at the transition to reheating (because super-horizon modes are frozen during matter and radiation dominance). The amplitude for the power spectrum should be the amplitude of these modes evaluated at the point inflation ends. Importantly, since there is such a strong time dependence in the power spectrum, changing the time inflation ends in this model by a small amount has a large effect on the amplitude of the power spectrum. 

This model, as with the large $\epsilon$ one, would give a totally wrong $n_s$ (in this case $n_s$ of exactly one), so this model can't describe our universe. Again, however, there could have been periods, especially early on, or following a sharp transition from a steeper to a flatter potential, when $\delta \approx 3$ for some time.  This situation is seen in two of the numerical examples below which do produce a growth of modes outside the horizon until slow roll is reached. 

In this case the power spectrum had to come out scale independent because there was essentially no scale to the potential. In more realistic models, scale dependence will appear both in the $\delta$ large, and slow roll regimes.  The fact that modes grow outside the horizon is independent of this and is a generic feature of large  $\delta$ evolution.





\section{Transient Periods of Fast Roll}

It's been pointed out \cite{Wang:1997cw} that in situations where not only the slow roll parameters are large, but also their derivatives are large, it becomes extremely difficult to find analytic solutions to the Mukhanov-Sasaki equation.  We found an analytic solution for the case of a $\phi^2$ potential as inflation is ending, but since the slow roll parameters are only large for about 1/2 an efolding before inflation ends, this case isn't particularly enlightening. Instead we consider numerically cases where the slow roll parameters become large earlier on during inflation, at observable scales, and then become small again. 

Situations where one starts the inflaton with extra kinetic energy have been considered in \cite{Contaldi:2003zv, Boyanovsky:2006pm, Nicholson:2007by, Lello:2013awa,Lello:2013mfa,Cicoli:2014bja,Handley:2014bqa,White:2014aua}. Others have considered possibilities where the shape of the potential changes such that it was steeper initially and then gets flatter to give suppression at low $l$, but without breaking slow roll throughout the observable part of the CMB spectrum, often considered in the context of tunneling \cite{Starobinsky:1992ts, Bousso:2013uia, Contaldi:2014zua, Bousso:2014jca, Hazra:2014jka, Hazra:2014goa}. The latter cases where one considers slow roll on a steeper potential and then slow roll on a more shallow potential are much easier to implement since slow roll approximations hold throughout,\footnote{ Except perhaps for a brief period immediately following the transition if the transition isn't smooth.} and because since inflation continues for some time, the Bunch-Davies initial conditions are valid. Both situations tend to give suppressed power at low $l$.

The question of BD initial conditions becomes an issue if one assumes inflation lasts `just long enough' such that the largest CMB scales represent the start of inflation, and some non-inflation period came before. We first consider numerically an arctan potential, this way there is a period of slow roll inflation where BD initial conditions are valid, and then a smooth transition to a period where the slow roll parameters become large, followed by another slow roll regime. Next we consider the familiar $\phi^2$ potential, but start with initial conditions such that $\phi$ starts with the maximum kinetic energy such that it's still inflating, 1/2 the potential. In this case, assuming a non-inflating period came before, BD initial conditions are at best a rough approximation which allows us to obtain a power spectrum that can be considered a lower bound.  This is because the BD initial conditions imply that the amplitude of the different modes has been reduced down to it's minimum allowed quantum value. Therefore, initial conditions set up by a non-inflating regime will likely give modes a larger initial amplitude.  For example, consider \cite{Leach:2000yw,Cicoli:2014bja} where they find that modes which are close to horizon size during a break in inflation are typically associated with enhanced power. 


\subsection{Arctan Potential}

\begin{figure}[H]
\centering
    \includegraphics[width=1.0\textwidth]{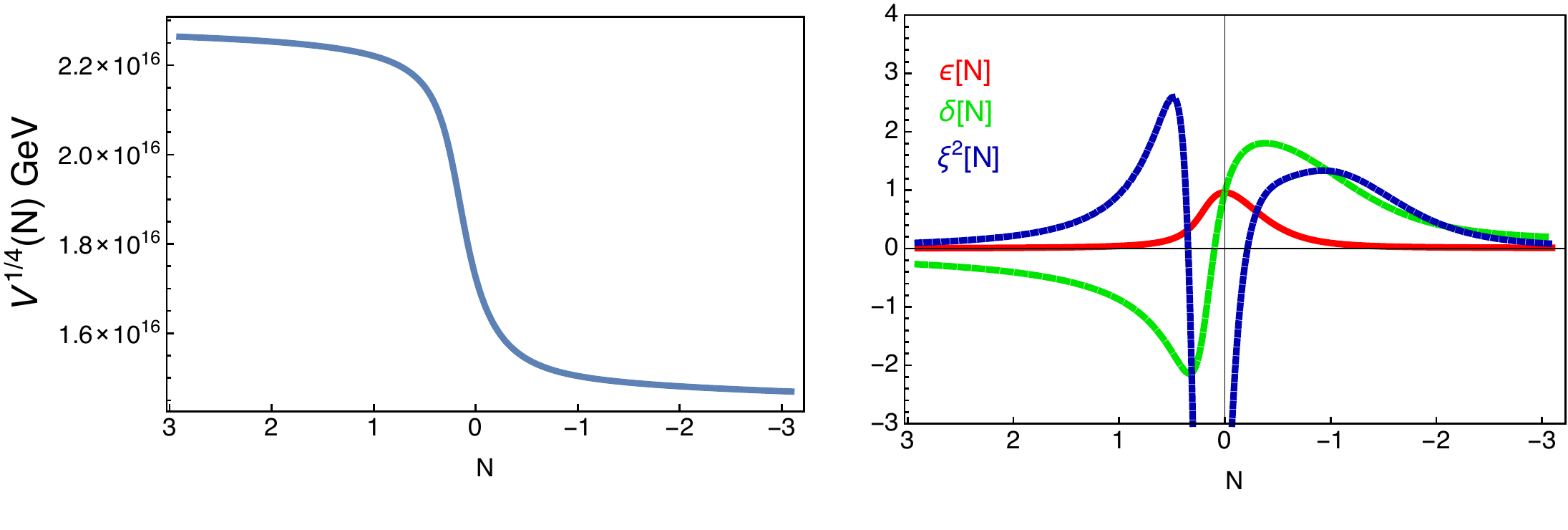}
    \caption{The figure depicts the potential and the three slow roll parameters, which appear in the Mukhanov-Sasaki equation for the arctan potential, plotted in the region where the potential becomes steep. $N=0$ is taken to be where $\epsilon$ is maximized.   \label{figa} }
\end{figure}

\begin{figure}[H]
\centering
    \includegraphics[width=1.0\textwidth]{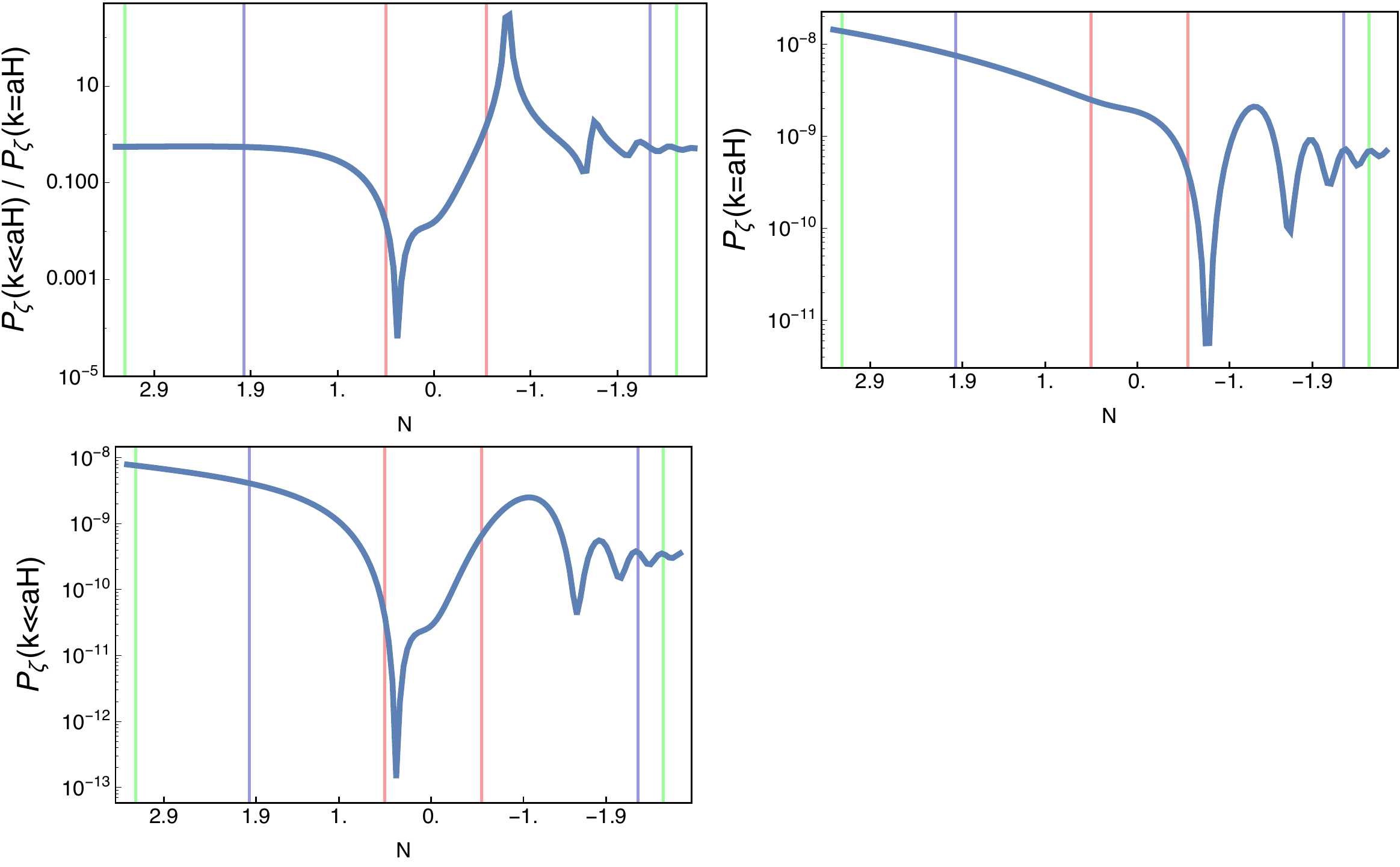}
    \caption{The top left figure depicts the ratio of the final power spectrum amplitude evaluated during the last slow roll regime to the power spectrum amplitude if evaluated at horizon crossing for the arctan potential. This shows how much modes either grow or decay while outside the horizon during the fast roll phase. The top right figure depicts the power spectrum amplitude evaluated at horizon crossing. The bottom left figure depicts the power spectrum amplitude evaluated during the final slow roll regime.  $N=0$ is taken to be where $\epsilon$ is maximized. The red vertical lines bracket the region $\epsilon \geq 0.25$, the green the region $\delta \geq 0.25$, and the blue the region $\xi^2 \geq 0.25$.  \label{fig1}}
\end{figure}

\begin{figure}[H]
\centering
    \includegraphics[width=.4\textwidth]{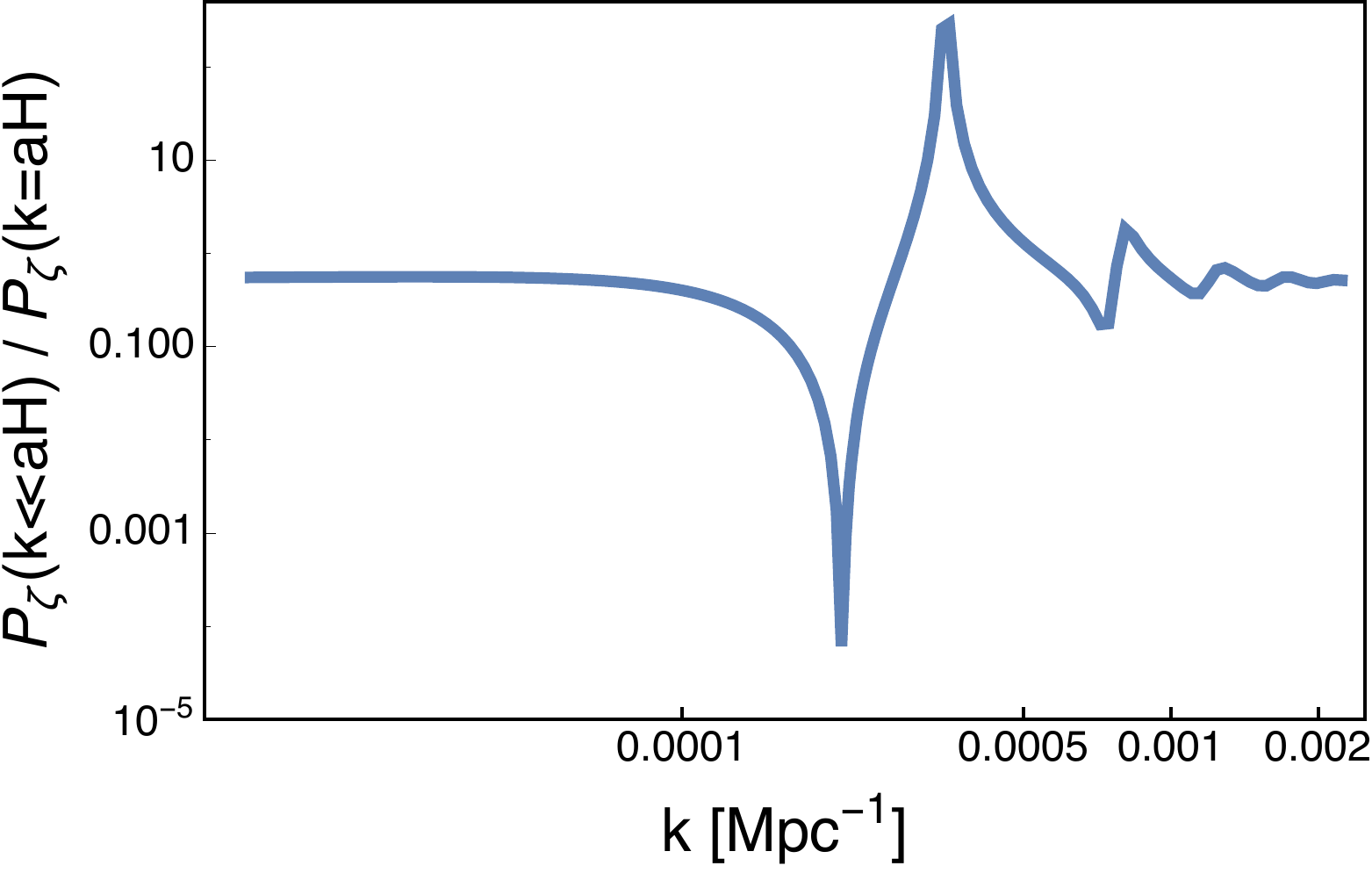}
    \caption{This figure shows the same results as the last figure but plotting in the more traditional way as a function of $k$.   \label{fig12}}
\end{figure}

For the arctan example we take:

 \begin{align}
V = \Lambda^4 \left(\arctan \left(- \frac{\phi}{\mu} \right) + C \right)
\end{align}

\noindent where the $+ C$ is added to make sure there is positive potential slow roll behavior after the steepening. One can choose to make the potential as steep as one wishes when $\phi =0$. We choose to keep $\epsilon < 1$ so inflation never actually stops. Figure \ref{figa} shows the potential and the evolution of the slow roll parameters around the transition.  $N$ stands for number of efolding where we fix $N=0$ to be the time where $\epsilon$ is maximized, and $N$ counts down such that larger $N$ means earlier time. Figure \ref{fig1} A shows the ratio of the power spectrum evaluated during the final slow roll regime to the power spectrum evaluated at horizon crossing in order to show the amount that modes either grow or decay outside the horizon during the fast roll regime. In plotting the power spectrum as a function of $N$, we are plotting different $k$ modes, with $N$ designating the horizon crossing time for each $k$ mode. The figure shows the same information as the more traditional power spectrum plots as a function of $k$, but with a redefined x-axis. We show the latter in Figure \ref{fig12} for reference, where the normalization of $k$ is arbitrary in this case, since this isn't meant to represent the actual physical power spectrum.  Plotting the power spectrum as a function of $N$ is meant to visually show how long the effects last.  Note in the early and late times in the figure, standard slow roll is taking place, and modes decay by about a factor of 2 outside the horizon. The vertical red lines bracket the region $\epsilon \geq 0.25$, the green the region $\delta \geq 0.25$, and the blue the region $\xi^2 \geq 0.25$. We find initially when $\epsilon$ is large, the modes decay outside the horizon, but as $\delta$ gets large the modes grow. There are also the oscillations/ wiggles one typically finds in the power spectrum following sharp transitions. 

Just tracing whether modes are growing or shrinking outside the horizon isn't enough to reproduce the full power spectrum. The amplitude of the modes as they reach horizon size is also changing in time, as depicted in Figure \ref{fig1} B. The modes that start with the smallest amplitude at horizon crossing subsequently grow the most outside the horizon. Incorporating both of these factors yields the power spectrum evaluated at late times, well into slow roll regime in Figure \ref{fig1} C.



\subsection{$\phi^2$ Potential with Fast Roll Initial Conditions}

In this example we take a standard $\phi^2$ potential but start the field off with extra kinetic energy such that there is a smooth transition from kinetic dominance, with energy loss due to Hubble friction, transitioning to fast roll inflation as $\epsilon$ drops to 1, and then eventually slow roll inflation as the kinetic energy continues to diminish and the attractor solution is reached. In this way the background equations are well defined. The difficulty comes in choosing initial conditions for the metric perturbations for the modes that are leaving the horizon shortly after inflation starts. Here as we have described, BD initial conditions are not necessarily valid.  Nevertheless, approximating the power spectrum using BD initial conditions should yield an underestimate for the final amplitude. Thus the plots we display should be considered as a minimum for the power spectrum from this fast roll period. 

\begin{figure}[H]
\centering
    \includegraphics[width=.6\textwidth]{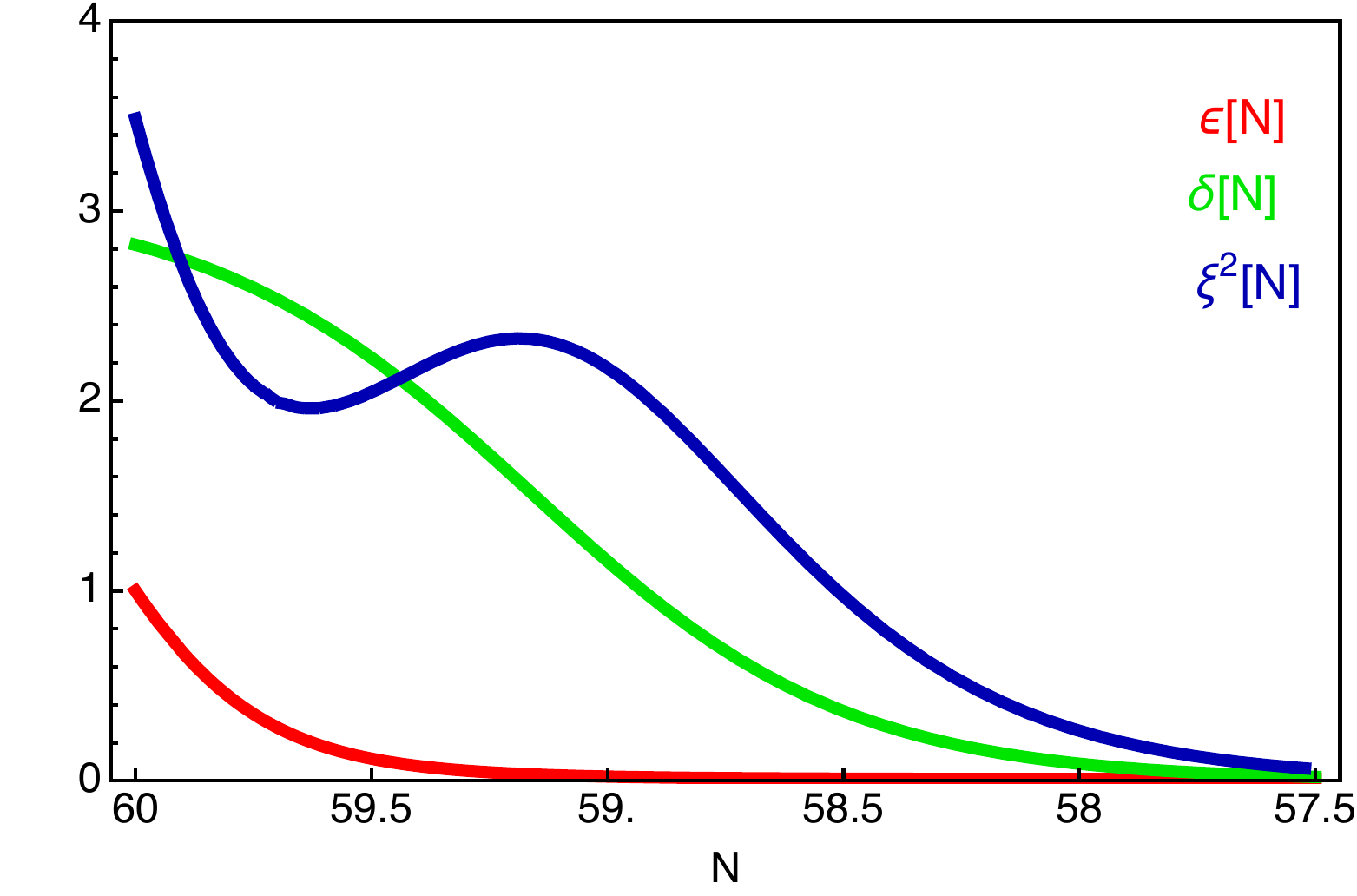}
    \caption{The figure depicts the three slow roll parameters which appear in the Mukhanov-Sasaki equation for the $\phi^2$ potential, plotted in the region where the potential is steep. $N=60$ corresponds to the onset of inflation with $N$ counting down. \label{figb}}
\end{figure}

\begin{figure}[H]
\centering
    \includegraphics[width=1.0\textwidth]{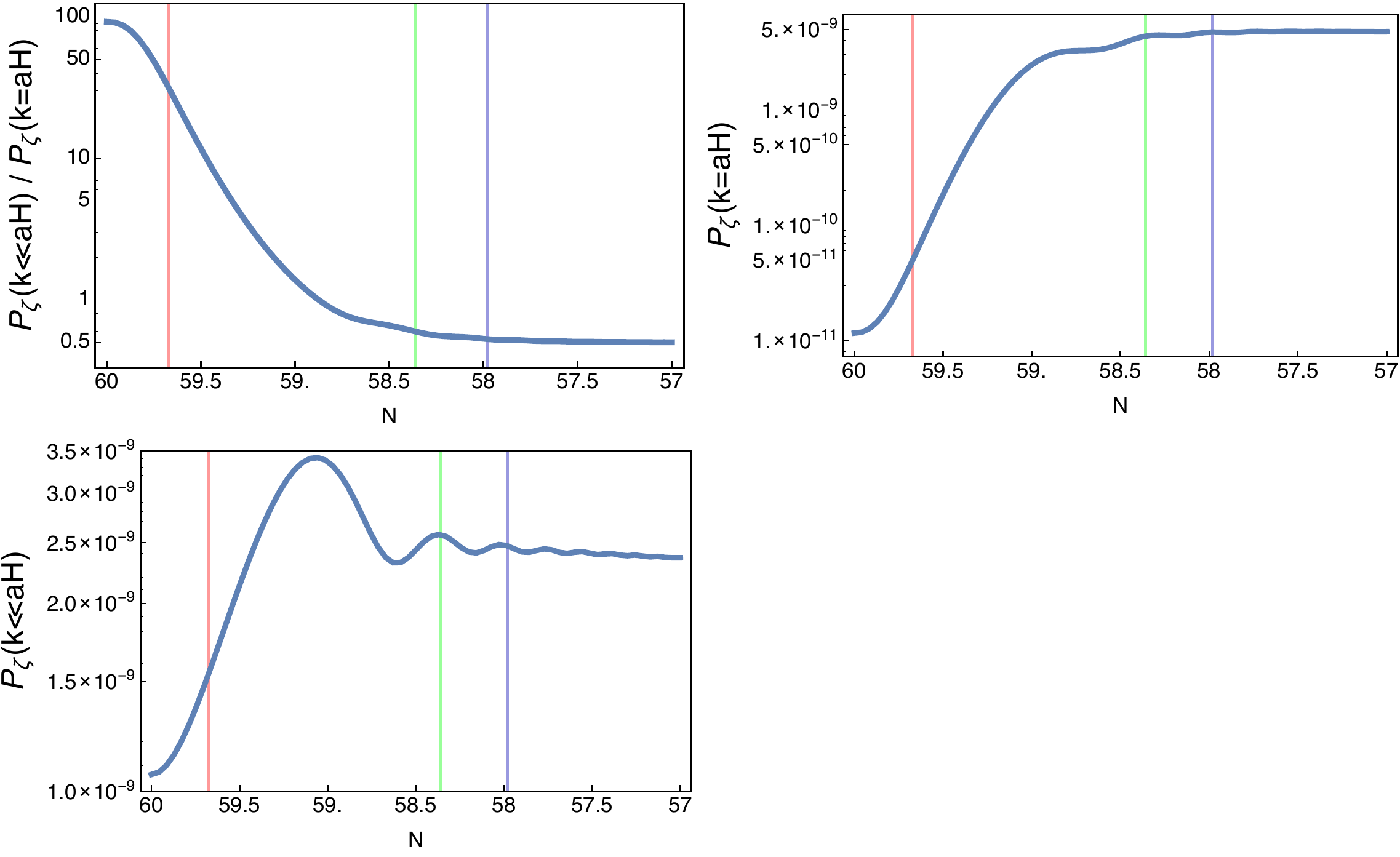}
    \caption{The top left figure depicts the ratio of the final power spectrum amplitude evaluated during the following slow roll regime to the power spectrum amplitude if evaluated at horizon crossing for the $\phi^2$ potential. This shows how much modes either grow or decay while outside the horizon during the fast roll phase. The top right figure depicts the power spectrum amplitude evaluated at horizon crossing. The bottom left figure depicts the final power spectrum evaluated during the slow roll regime. The red vertical line marks the region $\epsilon \geq 0.25$, the green the region $\delta \geq 0.25$, and the blue the region $\xi^2 \geq 0.25$.  \label{figc}}
\end{figure}

\begin{figure}[H]
\centering
    \includegraphics[width=0.4\textwidth]{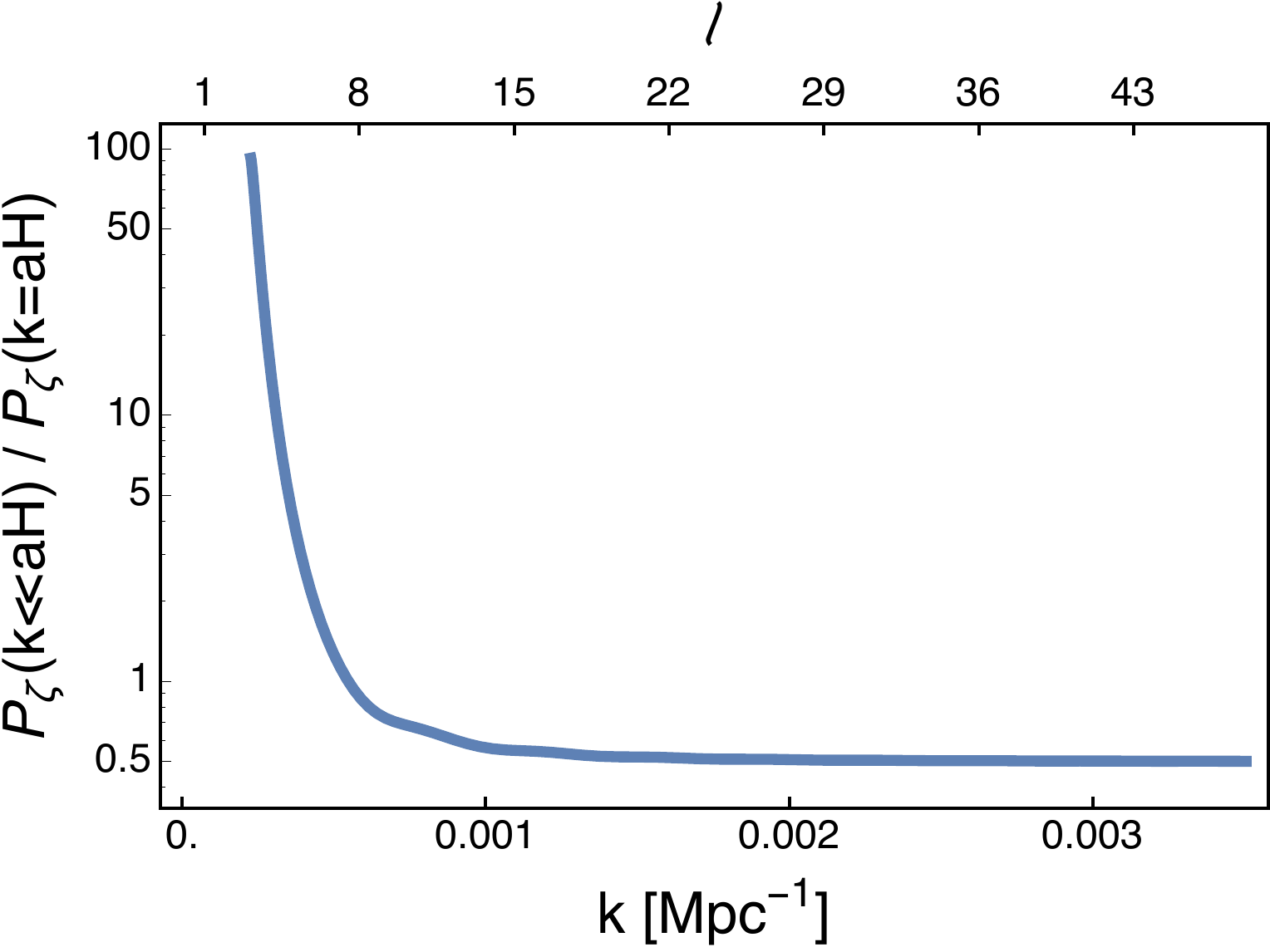}
    \caption{This figure shows the same results as the last figure but plotting in the more traditional way as a function of $k$. Note the $l$ axis is approximate as $P_{\zeta}$ for each $l$ is really an integral over $k$. Also we assume the first modes to freeze out with the onset of inflation correspond to the largest observable scales.  \label{figc2}}
\end{figure}

In Figure \ref{figb} we show the evolution of the three slow roll parameters appearing in the Mukhanov-Sasaki equation in the fast roll regime, where we define $N=60$ as the start of inflation when $\epsilon = 1$, and define $N$ so it's counting down towards the end of inflation. In Figure \ref{figc} A we show the ratio of the power spectrum evaluated in the late time limit, after slow roll has been established, to the power spectrum evaluated at horizon crossing, to show the evolution of the modes outside the horizon. Again, instead of displaying $k$ on the x-axis, we show $N$ as the time each $k$ mode reached horizon size, to visually show in efolds how long the effects of the fast roll initial conditions last. We show the more traditional power spectrum in Figure \ref{figc2} for reference.  In the slow roll limit, the familiar factor of 2 drop in the power spectrum after a mode leaves the horizon is restored, but initially we find modes grow outside the horizon as for the $\delta$ large example above. In Figure \ref{figc} B we show the power spectrum at horizon crossing and in Figure \ref{figc} C, the power spectrum in the late time limit for reference.  We find that although modes initially grow outside the horizon, they start from such a suppressed initial value that the final power spectrum for these initial modes is still smaller than for modes that exit the horizon during slow roll. 

\subsection{$R^2$ Potential with Fast Roll Initial Conditions}

\begin{figure}[H]
\centering
    \includegraphics[width=.5\textwidth]{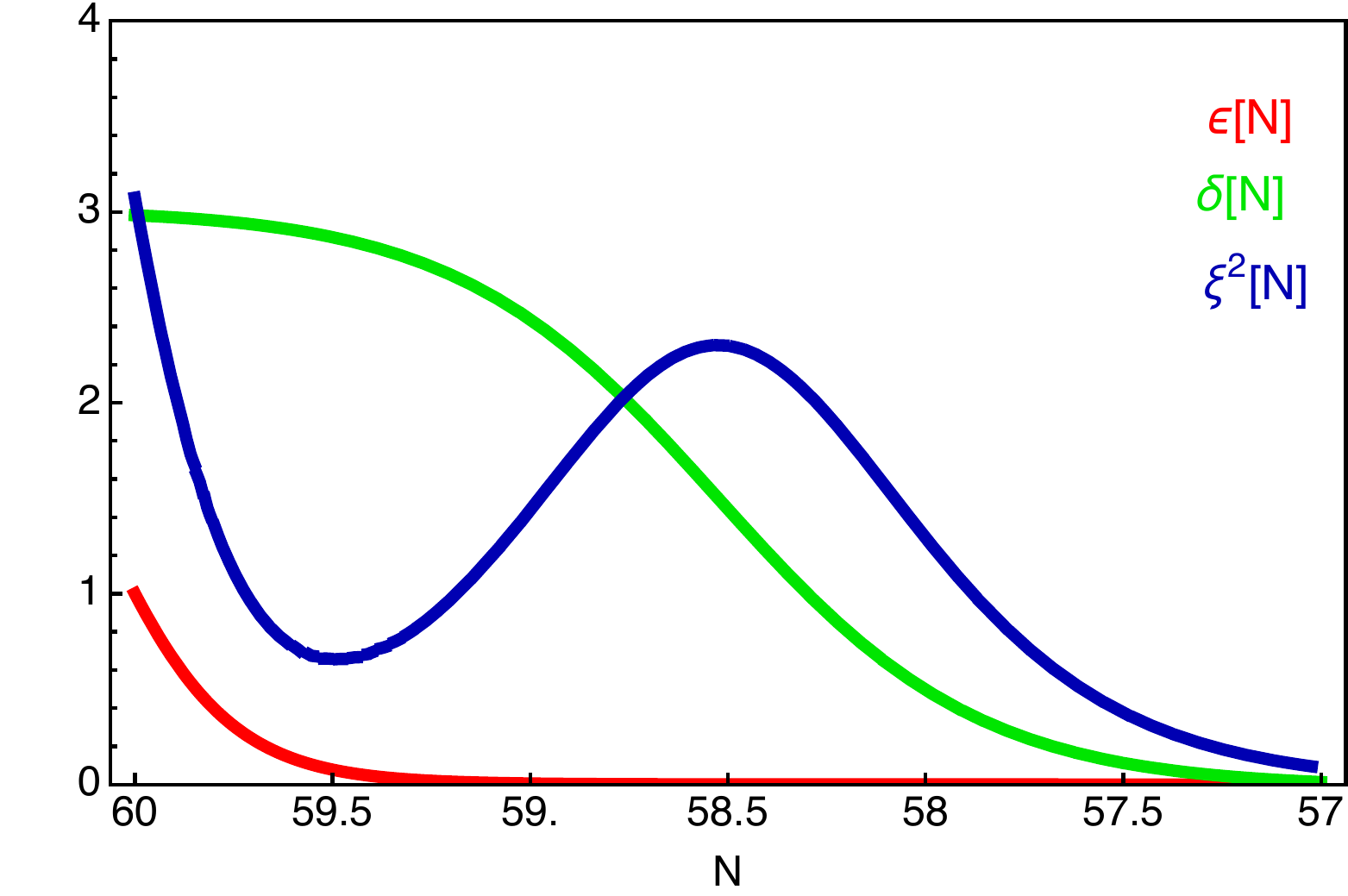}
    \caption{The figure depicts the three slow roll parameters which appear in the Mukhanov-Sasaki equation for the $R^2$ potential, plotted in the region where the potential is steep. $N=60$ corresponds to the onset of inflation with $N$ counting down.  \label{figd}}
\end{figure}

\begin{figure}[H]
\centering
    \includegraphics[width=1.0\textwidth]{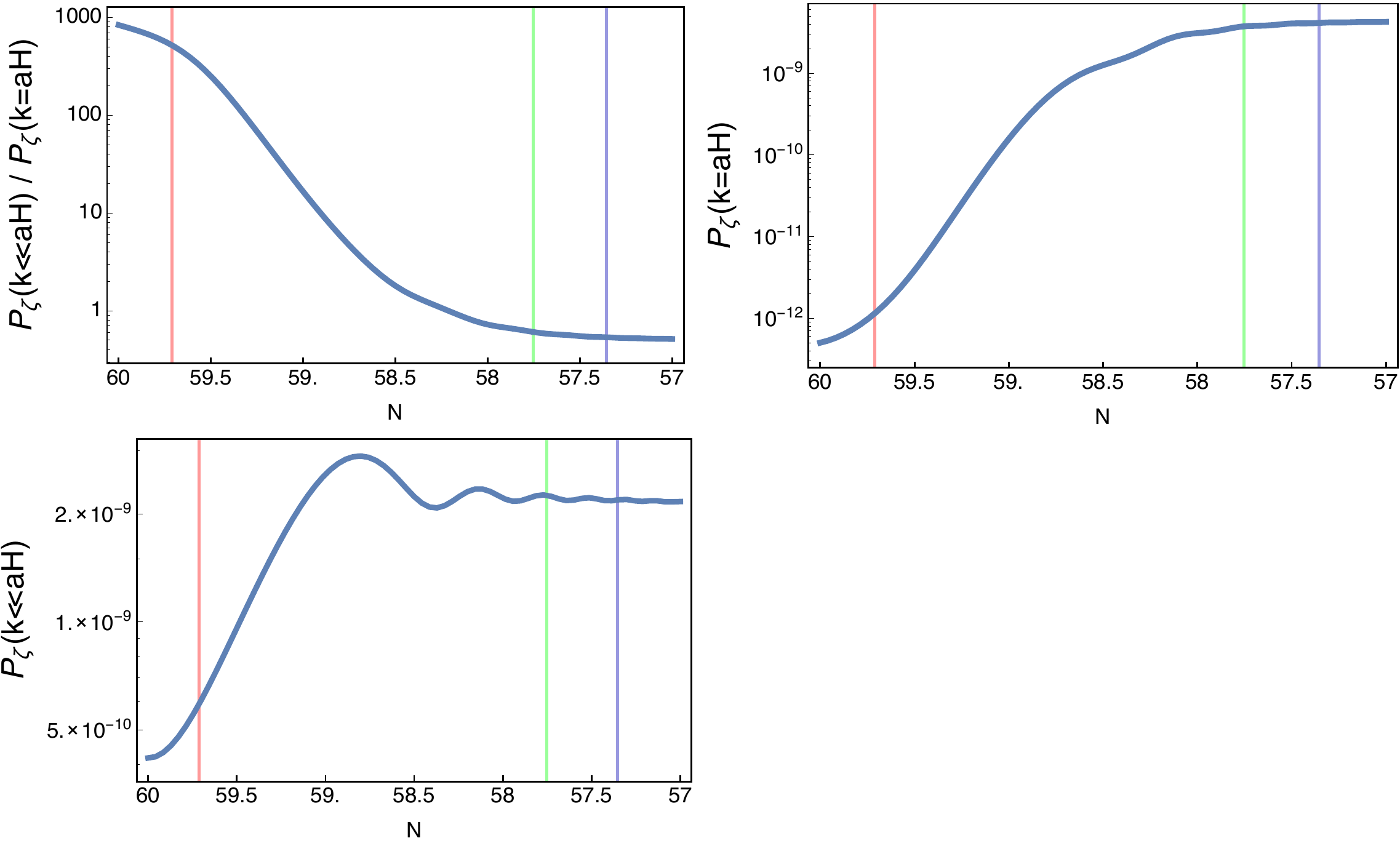}
    \caption{The top left figure depicts the ratio of the final power spectrum amplitude evaluated during the following slow roll regime to the power spectrum amplitude if evaluated at horizon crossing for the $R^2$ potential. This shows how much modes either grow or decay while outside the horizon during the fast roll phase. The top right figure depicts the power spectrum amplitude evaluated at horizon crossing. The bottom left figure depicts the final power spectrum amplitude evaluated during the slow roll regime. The red vertical line marks the region $\epsilon \geq 0.25$, the green the region $\delta \geq 0.25$, and the blue the region $\xi^2 \geq 0.25$.  \label{fige}}
\end{figure}

In this final numeric example we explore the effect of fast roll initial conditions for a flatter potential, the $R^2$ potential. We again give the field extra kinetic energy initially, and define $N=60$ as the start of inflation when $\epsilon = 1$. Figure \ref{figd} shows the evolution of the slow roll parameters for this case right around the fast roll transition. We find the flatter potential means that the $\delta \approx 3$ region lasts for a longer period of time. As mentioned above, this is because it takes longer for the $V_{\phi}$ term in the equation of motion to become relevant. Figure \ref{fige} shows the resultant power spectrum, where Figure \ref{fige} A shows the ratio of the late time to horizon crossing power spectra, displaying how much modes evolve outside the horizon in this case. Again, we find growth outside the horizon for a period until slow roll is reached. 

\begin{figure}[H]
\centering
    \includegraphics[width=.4\textwidth]{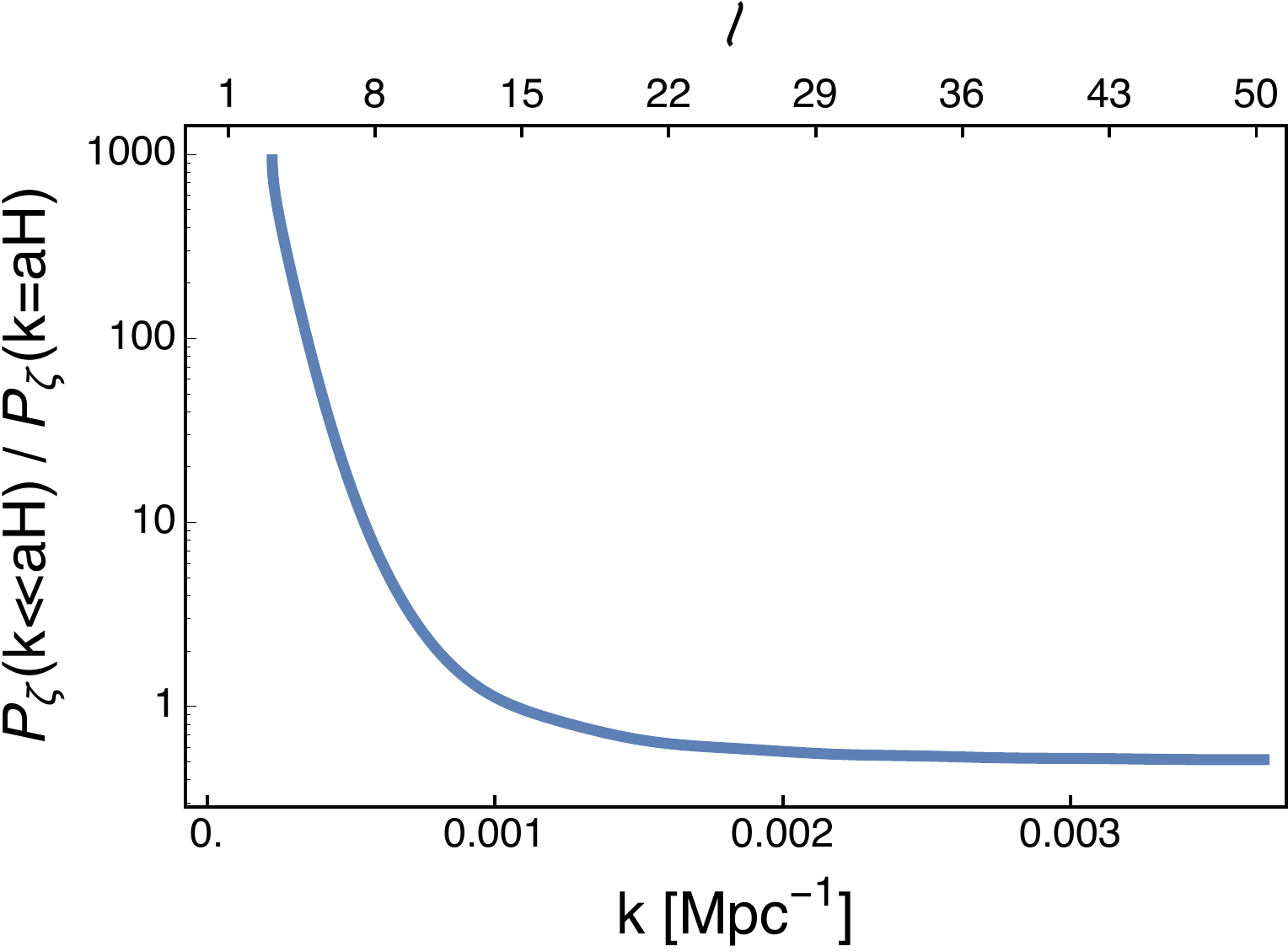}
    \caption{This figure shows the same results as the last figure but plotting in the more traditional way as a function of $k$. Note the $l$ axis is approximate as $P_{\zeta}$ for each $l$ is really an integral over $k$. Also we assume the first modes to freeze out with the onset of inflation correspond to the largest observable scales.    \label{fige2}}
\end{figure}

\section{Analytic Approximation}

Others \cite{Contaldi:2003zv, Cicoli:2014bja} have considered an analytic approximation where one takes a de Sitter solution in one region, a kinetic dominated solution in another range, and then just matches boundary conditions for $\mathcal{H}$, $\tau$, $u$, and $u_{\tau}$  between the regions where $u$ is the MS mode function.  As we show below, this method approximates well the late time suppression of the power spectrum, but doesn't reproduce the behavior of the modes outside the horizon. 

First during the kinetic dominated period, one takes the equation of motion and Friedmann equations without the potential terms:

 \begin{align}
\phi_{\tau \tau} + 2 \mathcal{H} \phi_{\tau}  = 0
 \end{align} 
 
 \noindent and
 
 \begin{align}
\mathcal{H}^2  = \frac{ \phi_{\tau}^2}{6 M_P^2} \, .
 \end{align} 
 
During kinetic dominance, $\epsilon \approx \delta \approx 3$ and $\xi^2 \approx 9$ and all three slow roll parameter's time derivatives are negligible. The general MS equation (eq. \ref{eq1}) then reduces to:

  \begin{align}
u_{\tau \tau} + u (k^2 +\frac{1}{4 \tau^2} ) = 0 \, .
 \end{align} 
 
\noindent This gives the standard slow roll solution but now with $\nu = 0$. If we wish to match onto BD initial conditions, this gives:
 
  \begin{align}
u_{KD} = \frac{\sqrt{\pi}}{2} \sqrt{- \tau} H_0^{(1)}(- k \tau) \, .
 \end{align} 
 
\begin{figure}[H]
\centering
    \includegraphics[width=1.0\textwidth]{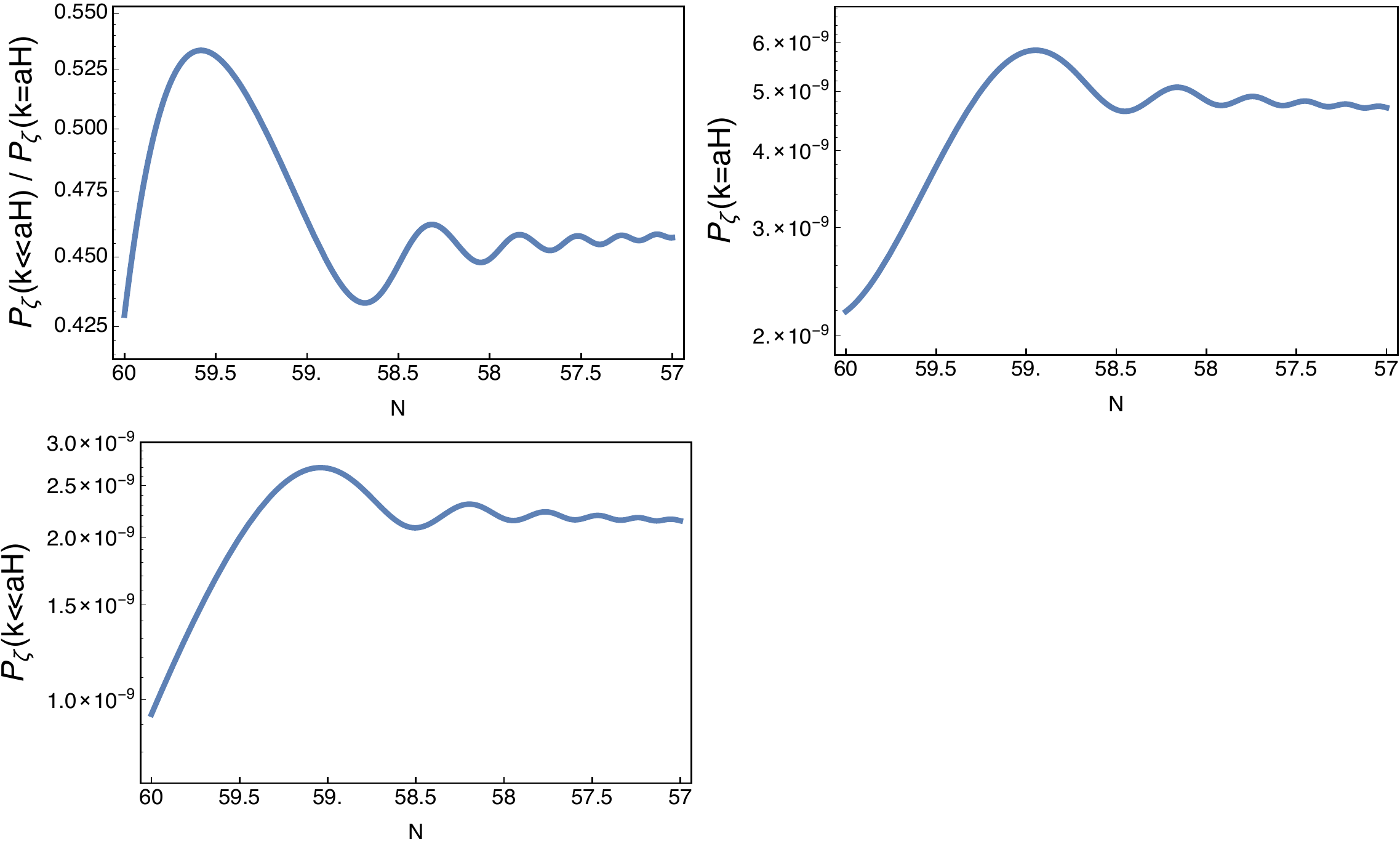}
    \caption{  This figure shows power spectra results for a kinetic dominated phase immediately transitioning to slow roll using the $\phi^2$ potential. The top left figure shows the ratio of the final amplitude to the horizon crossing amplitude. The top right figure shows the horizon crossing amplitude, and the bottom figure shows the late time amplitude.  \label{fig5}}
\end{figure}

 Next we take the traditional slow roll equations, making sure $\mathcal{H}$, $u$ and $u_{\tau}$ are continuous across the transition. Figure \ref{fig5} shows the results where we used slow roll parameters from the $\phi^2$ potential.

This simple analytic approach does well approximate the final power spectrum amplitude when evaluated at late times, almost exactly replicating the power spectrum for the $\phi^2$ potential as can be seen by comparing Figure \ref{fig5} to Figure \ref{figc}, but the approach doesn't well characterize the evolution of the modes as they are passing outside the horizon.  

\begin{figure}[H]
\centering
    \includegraphics[width=.4\textwidth]{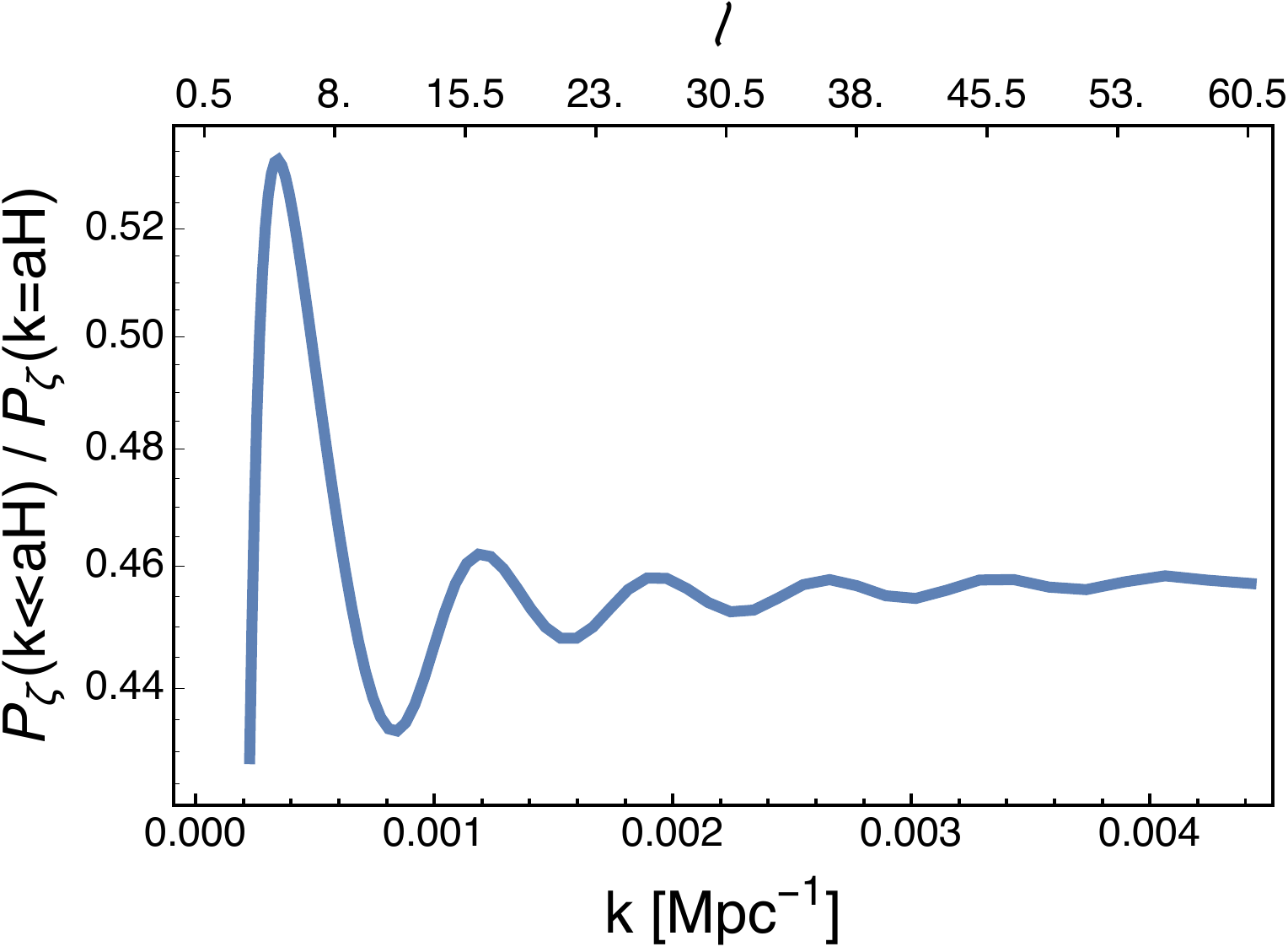}
    \caption{  This figure shows the same results as the last figure but plotting in the more trandiational way as a function of $k$. Note the $l$ axis is approximate as $P_{\zeta}$ for each $l$ is really an integral over $k$. Also we assume the first modes to freeze out with the onset of inflation correspond to the largest observable scales.  \label{fig55}}
\end{figure}

\section{Conclusions}

We investigate the growth of the comoving metric perturbations outside the horizon in two simple test cases where the slow roll parameters $\epsilon$ or $\delta$ become large. In the first case where $\epsilon$ becomes large, freeze out eventually occurs, but takes longer the closer $\epsilon$ is to 1. During the whole period that modes are evolving outside the horizon, they decay exponentially in efolding time. In the second case, when $\delta \approx 3$, freeze out doesn't occur until inflation ends, and the modes instead grow exponentially in efolding time outside the horizon. Neither case can provide the correct $n_s$ and so by itself can't describe what actually occurred during inflation. However, it is possible that during inflation there could be transient periods where the slow roll parameters become temporarily large.  In these cases one has to be careful to evaluate modes after slow roll has been reached rather then at horizon crossing, as we have shown. 

It is also worth noting that there could even be hints of such transient behavior in the CMB data, given the anomalously low power at low $l$ in the scalar power spectrum. We show that if one uses BD initial conditions and starts inflation with maximum kinetic energy (1/2 the potential), then one obtains a power spectrum that is suppressed at horizon crossing. The modes grow outside the horizon, but the final power spectrum is still suppressed for those initial modes, with the effect lasting for a few efolds. The growth in the modes outside the horizon lasts until $\delta$ returns to its slow roll value, which takes longer for flatter potentials.

Such cases, where the initial suppression at low $l$ can be linked to the earliest stages of inflation, are therefore phenomenologically interesting, but there is an open question as to what to use for initial conditions, which is therefore an important issue when trying to compare to actual data. We used BD initial conditions in this case for convenience, but as we have noted, this should only be viewed as a lower bound. 

For the one transient case where initial conditions are unambiguous (an arctan potential), where there is an initial and final slow roll regime, and the slow roll parameters get large in-between, we find suppression when $\epsilon$ is large, followed by oscillations, typical when there are sudden transitions in power spectra. \\



\acknowledgments

It is a pleasure to thank Will Kinney for useful correspondence. This research was supported by a grant from the DOE.

\bibliographystyle{JHEPmodplain}
\bibliography{references}

\providecommand{\href}[2]{#2}\begingroup\raggedright\begin{thebibliography}{10}

\bibitem{Takahashi:2013tj}
F.~Takahashi, {\it {The Spectral Index and its Running in Axionic Curvaton}},
  {\sl JCAP} {\bf 1306} (2013) 013, [\href{http://arxiv.org/abs/1301.2834}{{\sf
  arXiv:1301.2834}}],
  [\href{http://dx.doi.org/10.1088/1475-7516/2013/06/013}{{\sf
  doi:10.1088/1475-7516/2013/06/013}}].

\bibitem{Hou:2012xq}
Z.~Hou, C.~Reichardt, K.~Story, B.~Follin, R.~Keisler, {\em et~al.}, {\it
  {Constraints on Cosmology from the Cosmic Microwave Background Power Spectrum
  of the 2500 deg$^2$ SPT-SZ Survey}},  {\sl Astrophys.J.} {\bf 782} (2014) 74,
  [\href{http://arxiv.org/abs/1212.6267}{{\sf arXiv:1212.6267}}],
  [\href{http://dx.doi.org/10.1088/0004-637X/782/2/74}{{\sf
  doi:10.1088/0004-637X/782/2/74}}].

\bibitem{Ade:2015lrj}
{\bf Planck} Collaboration, P.~Ade {\em et~al.}, {\it {Planck 2015 results. XX.
  Constraints on inflation}},  \href{http://arxiv.org/abs/1502.02114}{{\sf
  arXiv:1502.02114}}.

\bibitem{Ade:2013kta}
{\bf Planck} Collaboration, P.~Ade {\em et~al.}, {\it {Planck 2013 results. XV.
  CMB power spectra and likelihood}},  {\sl Astron.Astrophys.} {\bf 571} (2014)
  A15, [\href{http://arxiv.org/abs/1303.5075}{{\sf arXiv:1303.5075}}],
  [\href{http://dx.doi.org/10.1051/0004-6361/201321573}{{\sf
  doi:10.1051/0004-6361/201321573}}].

\bibitem{Bennett:1996ce}
C.~Bennett, A.~Banday, K.~Gorski, G.~Hinshaw, P.~Jackson, {\em et~al.}, {\it
  {Four year COBE DMR cosmic microwave background observations: Maps and basic
  results}},  {\sl Astrophys.J.} {\bf 464} (1996) L1--L4,
  [\href{http://arxiv.org/abs/astro-ph/9601067}{{\sf arXiv:astro-ph/9601067}}],
  [\href{http://dx.doi.org/10.1086/310075}{{\sf doi:10.1086/310075}}].

\bibitem{Bennett:2003bz}
{\bf WMAP} Collaboration, C.~Bennett {\em et~al.}, {\it {First year Wilkinson
  Microwave Anisotropy Probe (WMAP) observations: Preliminary maps and basic
  results}},  {\sl Astrophys.J.Suppl.} {\bf 148} (2003) 1--27,
  [\href{http://arxiv.org/abs/astro-ph/0302207}{{\sf arXiv:astro-ph/0302207}}],
  [\href{http://dx.doi.org/10.1086/377253}{{\sf doi:10.1086/377253}}].

\bibitem{Spergel:2003cb}
{\bf WMAP} Collaboration, D.~Spergel {\em et~al.}, {\it {First year Wilkinson
  Microwave Anisotropy Probe (WMAP) observations: Determination of cosmological
  parameters}},  {\sl Astrophys.J.Suppl.} {\bf 148} (2003) 175--194,
  [\href{http://arxiv.org/abs/astro-ph/0302209}{{\sf arXiv:astro-ph/0302209}}],
  [\href{http://dx.doi.org/10.1086/377226}{{\sf doi:10.1086/377226}}].

\bibitem{White:1993jr}
M.~J. White, L.~M. Krauss, and J.~Silk, {\it {Inflation and the statistics of
  cosmic microwave background anisotropies: From 1-degree to COBE}},  {\sl
  Astrophys. J.} {\bf 418} (1993) 535,
  [\href{http://arxiv.org/abs/astro-ph/9303009}{{\sf arXiv:astro-ph/9303009}}],
  [\href{http://dx.doi.org/10.1086/173415}{{\sf doi:10.1086/173415}}].

\bibitem{Contaldi:2003zv}
C.~R. Contaldi, M.~Peloso, L.~Kofman, and A.~D. Linde, {\it {Suppressing the
  lower multipoles in the CMB anisotropies}},  {\sl JCAP} {\bf 0307} (2003)
  002, [\href{http://arxiv.org/abs/astro-ph/0303636}{{\sf
  arXiv:astro-ph/0303636}}],
  [\href{http://dx.doi.org/10.1088/1475-7516/2003/07/002}{{\sf
  doi:10.1088/1475-7516/2003/07/002}}].

\bibitem{Liddle:1994dx}
A.~R. Liddle, P.~Parsons, and J.~D. Barrow, {\it {Formalizing the slow roll
  approximation in inflation}},  {\sl Phys.Rev.} {\bf D50} (1994) 7222--7232,
  [\href{http://arxiv.org/abs/astro-ph/9408015}{{\sf arXiv:astro-ph/9408015}}],
  [\href{http://dx.doi.org/10.1103/PhysRevD.50.7222}{{\sf
  doi:10.1103/PhysRevD.50.7222}}].

\bibitem{Tsamis:2003px}
N.~Tsamis and R.~P. Woodard, {\it {Improved estimates of cosmological
  perturbations}},  {\sl Phys.Rev.} {\bf D69} (2004) 084005,
  [\href{http://arxiv.org/abs/astro-ph/0307463}{{\sf arXiv:astro-ph/0307463}}],
  [\href{http://dx.doi.org/10.1103/PhysRevD.69.084005}{{\sf
  doi:10.1103/PhysRevD.69.084005}}].

\bibitem{Kinney:2005vj}
W.~H. Kinney, {\it {Horizon crossing and inflation with large eta}},  {\sl
  Phys.Rev.} {\bf D72} (2005) 023515,
  [\href{http://arxiv.org/abs/gr-qc/0503017}{{\sf arXiv:gr-qc/0503017}}],
  [\href{http://dx.doi.org/10.1103/PhysRevD.72.023515}{{\sf
  doi:10.1103/PhysRevD.72.023515}}].

\bibitem{Tzirakis:2007bf}
K.~Tzirakis and W.~H. Kinney, {\it {Inflation over the hill}},  {\sl Phys.Rev.}
  {\bf D75} (2007) 123510, [\href{http://arxiv.org/abs/astro-ph/0701432}{{\sf
  arXiv:astro-ph/0701432}}],
  [\href{http://dx.doi.org/10.1103/PhysRevD.75.123510}{{\sf
  doi:10.1103/PhysRevD.75.123510}}].

\bibitem{Namjoo:2012aa}
M.~H. Namjoo, H.~Firouzjahi, and M.~Sasaki, {\it {Violation of non-Gaussianity
  consistency relation in a single field inflationary model}},  {\sl
  Europhys.Lett.} {\bf 101} (2013) 39001,
  [\href{http://arxiv.org/abs/1210.3692}{{\sf arXiv:1210.3692}}],
  [\href{http://dx.doi.org/10.1209/0295-5075/101/39001}{{\sf
  doi:10.1209/0295-5075/101/39001}}].

\bibitem{Martin:2012pe}
J.~Martin, H.~Motohashi, and T.~Suyama, {\it {Ultra Slow-Roll Inflation and the
  non-Gaussianity Consistency Relation}},  {\sl Phys.Rev.} {\bf D87} (2013),
  no.~2 023514, [\href{http://arxiv.org/abs/1211.0083}{{\sf arXiv:1211.0083}}],
  [\href{http://dx.doi.org/10.1103/PhysRevD.87.023514}{{\sf
  doi:10.1103/PhysRevD.87.023514}}].

\bibitem{Motohashi:2014ppa}
H.~Motohashi, A.~A. Starobinsky, and J.~Yokoyama, {\it {Inflation with a
  constant rate of roll}},  \href{http://arxiv.org/abs/1411.5021}{{\sf
  arXiv:1411.5021}}.

\bibitem{Mooij:2015yka}
S.~Mooij, G.~A. Palma, and A.~E. Romano, {\it {Consistently violating the
  non-Gaussian consistency relation}},
  \href{http://arxiv.org/abs/1502.03458}{{\sf arXiv:1502.03458}}.

\bibitem{Wang:1997cw}
L.-M. Wang, V.~F. Mukhanov, and P.~J. Steinhardt, {\it {On the problem of
  predicting inflationary perturbations}},  {\sl Phys.Lett.} {\bf B414} (1997)
  18--27, [\href{http://arxiv.org/abs/astro-ph/9709032}{{\sf
  arXiv:astro-ph/9709032}}],
  [\href{http://dx.doi.org/10.1016/S0370-2693(97)01166-0}{{\sf
  doi:10.1016/S0370-2693(97)01166-0}}].

\bibitem{Boyanovsky:2006pm}
D.~Boyanovsky, H.~J. de~Vega, and N.~Sanchez, {\it {CMB quadrupole suppression.
  2. The early fast roll stage}},  {\sl Phys.Rev.} {\bf D74} (2006) 123007,
  [\href{http://arxiv.org/abs/astro-ph/0607487}{{\sf arXiv:astro-ph/0607487}}],
  [\href{http://dx.doi.org/10.1103/PhysRevD.74.123007}{{\sf
  doi:10.1103/PhysRevD.74.123007}}].

\bibitem{Nicholson:2007by}
G.~Nicholson and C.~R. Contaldi, {\it {The large scale CMB cut-off and the
  tensor-to-scalar ratio}},  {\sl JCAP} {\bf 0801} (2008) 002,
  [\href{http://arxiv.org/abs/astro-ph/0701783}{{\sf arXiv:astro-ph/0701783}}],
  [\href{http://dx.doi.org/10.1088/1475-7516/2008/01/002}{{\sf
  doi:10.1088/1475-7516/2008/01/002}}].

\bibitem{Lello:2013awa}
L.~Lello and D.~Boyanovsky, {\it {Tensor to scalar ratio and large scale power
  suppression from pre-slow roll initial conditions}},  {\sl JCAP} {\bf 1405}
  (2014) 029, [\href{http://arxiv.org/abs/1312.4251}{{\sf arXiv:1312.4251}}],
  [\href{http://dx.doi.org/10.1088/1475-7516/2014/05/029}{{\sf
  doi:10.1088/1475-7516/2014/05/029}}].

\bibitem{Lello:2013mfa}
L.~Lello, D.~Boyanovsky, and R.~Holman, {\it {Pre-slow roll initial conditions:
  large scale power suppression and infrared aspects during inflation}},  {\sl
  Phys.Rev.} {\bf D89} (2014), no.~6 063533,
  [\href{http://arxiv.org/abs/1307.4066}{{\sf arXiv:1307.4066}}],
  [\href{http://dx.doi.org/10.1103/PhysRevD.89.063533}{{\sf
  doi:10.1103/PhysRevD.89.063533}}].

\bibitem{Cicoli:2014bja}
M.~Cicoli, S.~Downes, B.~Dutta, F.~G. Pedro, and A.~Westphal, {\it {Just enough
  inflation: power spectrum modifications at large scales}},  {\sl JCAP} {\bf
  1412} (2014), no.~12 030, [\href{http://arxiv.org/abs/1407.1048}{{\sf
  arXiv:1407.1048}}],
  [\href{http://dx.doi.org/10.1088/1475-7516/2014/12/030}{{\sf
  doi:10.1088/1475-7516/2014/12/030}}].

\bibitem{Handley:2014bqa}
W.~Handley, S.~Brechet, A.~Lasenby, and M.~Hobson, {\it {Kinetic Initial
  Conditions for Inflation}},  {\sl Phys.Rev.} {\bf D89} (2014), no.~6 063505,
  [\href{http://arxiv.org/abs/1401.2253}{{\sf arXiv:1401.2253}}],
  [\href{http://dx.doi.org/10.1103/PhysRevD.89.063505}{{\sf
  doi:10.1103/PhysRevD.89.063505}}].

\bibitem{White:2014aua}
J.~White, Y.-l. Zhang, and M.~Sasaki, {\it {Scalar suppression on large scales
  in open inflation}},  {\sl Phys. Rev.} {\bf D90} (2014), no.~8 083517,
  [\href{http://arxiv.org/abs/1407.5816}{{\sf arXiv:1407.5816}}],
  [\href{http://dx.doi.org/10.1103/PhysRevD.90.083517}{{\sf
  doi:10.1103/PhysRevD.90.083517}}].

\bibitem{Starobinsky:1992ts}
A.~A. Starobinsky, {\it {Spectrum of adiabatic perturbations in the universe
  when there are singularities in the inflation potential}},  {\sl JETP Lett.}
  {\bf 55} (1992) 489--494.

\bibitem{Bousso:2013uia}
R.~Bousso, D.~Harlow, and L.~Senatore, {\it {Inflation after False Vacuum
  Decay: Observational Prospects after Planck}},  {\sl Phys.Rev.} {\bf D91}
  (2015), no.~8 083527, [\href{http://arxiv.org/abs/1309.4060}{{\sf
  arXiv:1309.4060}}], [\href{http://dx.doi.org/10.1103/PhysRevD.91.083527}{{\sf
  doi:10.1103/PhysRevD.91.083527}}].

\bibitem{Contaldi:2014zua}
C.~R. Contaldi, M.~Peloso, and L.~Sorbo, {\it {Suppressing the impact of a high
  tensor-to-scalar ratio on the temperature anisotropies}},  {\sl JCAP} {\bf
  1407} (2014) 014, [\href{http://arxiv.org/abs/1403.4596}{{\sf
  arXiv:1403.4596}}],
  [\href{http://dx.doi.org/10.1088/1475-7516/2014/07/014}{{\sf
  doi:10.1088/1475-7516/2014/07/014}}].

\bibitem{Bousso:2014jca}
R.~Bousso, D.~Harlow, and L.~Senatore, {\it {Inflation After False Vacuum
  Decay: New Evidence from BICEP2}},  {\sl JCAP} {\bf 1412} (2014), no.~12 019,
  [\href{http://arxiv.org/abs/1404.2278}{{\sf arXiv:1404.2278}}],
  [\href{http://dx.doi.org/10.1088/1475-7516/2014/12/019}{{\sf
  doi:10.1088/1475-7516/2014/12/019}}].

\bibitem{Hazra:2014jka}
D.~K. Hazra, A.~Shafieloo, G.~F. Smoot, and A.~A. Starobinsky, {\it {Inflation
  with Whip-Shaped Suppressed Scalar Power Spectra}},  {\sl Phys.Rev.Lett.}
  {\bf 113} (2014), no.~7 071301, [\href{http://arxiv.org/abs/1404.0360}{{\sf
  arXiv:1404.0360}}],
  [\href{http://dx.doi.org/10.1103/PhysRevLett.113.071301}{{\sf
  doi:10.1103/PhysRevLett.113.071301}}].

\bibitem{Hazra:2014goa}
D.~K. Hazra, A.~Shafieloo, G.~F. Smoot, and A.~A. Starobinsky, {\it {Wiggly
  Whipped Inflation}},  {\sl JCAP} {\bf 1408} (2014) 048,
  [\href{http://arxiv.org/abs/1405.2012}{{\sf arXiv:1405.2012}}],
  [\href{http://dx.doi.org/10.1088/1475-7516/2014/08/048}{{\sf
  doi:10.1088/1475-7516/2014/08/048}}].

\bibitem{Leach:2000yw}
S.~M. Leach and A.~R. Liddle, {\it {Inflationary perturbations near horizon
  crossing}},  {\sl Phys. Rev.} {\bf D63} (2001) 043508,
  [\href{http://arxiv.org/abs/astro-ph/0010082}{{\sf arXiv:astro-ph/0010082}}],
  [\href{http://dx.doi.org/10.1103/PhysRevD.63.043508}{{\sf
  doi:10.1103/PhysRevD.63.043508}}].

\bibitem{Leach:2001zf}
S.~M. Leach, M.~Sasaki, D.~Wands, and A.~R. Liddle, {\it {Enhancement of
  superhorizon scale inflationary curvature perturbations}},  {\sl Phys. Rev.}
  {\bf D64} (2001) 023512, [\href{http://arxiv.org/abs/astro-ph/0101406}{{\sf
  arXiv:astro-ph/0101406}}],
  [\href{http://dx.doi.org/10.1103/PhysRevD.64.023512}{{\sf
  doi:10.1103/PhysRevD.64.023512}}].

\bibitem{Jain:2007au}
R.~K. Jain, P.~Chingangbam, and L.~Sriramkumar, {\it {On the evolution of
  tachyonic perturbations at super-Hubble scales}},  {\sl JCAP} {\bf 0710}
  (2007) 003, [\href{http://arxiv.org/abs/astro-ph/0703762}{{\sf
  arXiv:astro-ph/0703762}}],
  [\href{http://dx.doi.org/10.1088/1475-7516/2007/10/003}{{\sf
  doi:10.1088/1475-7516/2007/10/003}}].

\bibitem{Jain:2008dw}
R.~K. Jain, P.~Chingangbam, J.-O. Gong, L.~Sriramkumar, and T.~Souradeep, {\it
  {Punctuated inflation and the low CMB multipoles}},  {\sl JCAP} {\bf 0901}
  (2009) 009, [\href{http://arxiv.org/abs/0809.3915}{{\sf arXiv:0809.3915}}],
  [\href{http://dx.doi.org/10.1088/1475-7516/2009/01/009}{{\sf
  doi:10.1088/1475-7516/2009/01/009}}].

\bibitem{Jain:2009pm}
R.~K. Jain, P.~Chingangbam, L.~Sriramkumar, and T.~Souradeep, {\it {The
  tensor-to-scalar ratio in punctuated inflation}},  {\sl Phys. Rev.} {\bf D82}
  (2010) 023509, [\href{http://arxiv.org/abs/0904.2518}{{\sf
  arXiv:0904.2518}}], [\href{http://dx.doi.org/10.1103/PhysRevD.82.023509}{{\sf
  doi:10.1103/PhysRevD.82.023509}}].

\end{thebibliography}\endgroup

\end{document}